\documentclass[aps,twocolumn,pre]{revtex4}
\usepackage{graphicx}
\newcommand{\n}{\noindent}

\begin{document}

\title{Accelerator modes of square well system}
\author{R. Sankaranarayanan\footnote {Present Address: Centre for Nonlinear
Dynamics, Department of Physics, Bharathidasan University, Tiruchirappalli
620024, India} and V.B. Sheorey}
\affiliation{Physical Research Laboratory, Navrangpura, Ahmedabad 380009,
India.}

\begin{abstract}
We study accelerator modes of a particle, confined in an one-dimensional
infinite square well potential, subjected to a time-periodic pulsed field.
Dynamics of such a particle can be described by one generalization of the
kicked rotor. In comparison with the kicked rotor, this generalization is
shown to have a much larger parametric space for existence of the modes.
Using this freedom we provide evidence that accelerator mode assisted
anomalous transport is greatly enhanced when low order resonances are exposed
at the border of chaos. We also present signature of the enhanced transport
in the quantum domain. 
\end{abstract}
\maketitle

\section {Introduction}

In recent years there is a surge of interest in studying mixed systems, 
with both regular and chaotic regions coexisting in phase space. Mixed
systems are important as a majority of dynamical systems found in nature
fall in this category. One of the outstanding problems is to resolve how
chaotic orbits are influenced by regular regions in the long time limit. In
this direction of study, two dimensional area-preserving maps are promising
candidates as they exhibit many generic features. A text book paradigm of
this kind is the standard map of kicked rotor \cite{lich}. For certain
values of system parameter, this map possesses special kind of orbits
called {\it Accelerator Modes} (AM) \cite{chir}. If an initial state of the
rotor corresponds to the AM, then it is uniformly accelerated after each
kick. In other words, momentum of the rotor increases linearly with time.
This is to be contrasted with generic chaotic orbits that diffuse in
momentum, i.e., momentum variance of an ensemble of chaotic orbits increases
linearly in time leading to {\it normal diffusion}.

AM are seen in phase space as small regular islands embedded in a chaotic
sea. They are separated from chaotic region by cantori (partial-barriers). 
If a chaotic orbit enters through the cantori, it sticks to the cantori
for very long time and thus is accelerated along with the modes. Considering
$n$ as time, such a long excursion of a chaotic orbit is an example of {\it
L\'{e}vy flight} \cite{shle0}. L\'{e}vy flight, some times called as {\it super
diffusion}, is characterized by $\langle p_n^2\rangle \sim n^\gamma$ with
$\gamma>1$. AM assisted anomalous transport of this kind has been already
studied with different motivations \cite{ishi}. It should be noted that the
above mechanism of anomalous effect is numerically ascertained. 

Further studies have shown that these modes influence corresponding quantum
system as well \cite{hans,zas}. In particular, the existence of regular
islands in chaotic sea significantly alter the localization of quasienergy
states in angular momentum, leading to enhancement in quantum transport.
Developments on confining atoms in magneto-optical traps have opened
experimental feasibility for atom-optics realizations of the kicked
rotor. One such realization, in which an ensemble of cold cesium atoms are
exposed to a periodically pulsed standing wave light, has shown that for
certain values of the kick amplitude the momentum distribution of the atomic
sample is non-exponentially localized \cite{klap98}. This then is shown as
an effect of AM in the quantum system, in contrast to the generic exponential
localization.

Semiconductor based technological developments, on the other hand, have
shown that it is possible to fabricate potential wells on atomic scales.
Motion of electrons in such quantum wells in presence of an external
electromagnetic field has been studied for experimental signatures of
quantum chaos \cite{wilk}. A simple model of this kind is a particle
confined in a one-dimensional infinite square well exposed to a time periodic
pulse field. This system is known to be one generalization of the kicked
rotor \cite{sankar}. The virtue of this generalization arises from two
length scales of the system, namely, width of the well and field wavelength.
If the well width is an integer multiple of the field wavelength, kick to
kick dynamics of the particle is equivalent to that of the rotor. Otherwise,
the dynamics is described by a discontinuous map resulting in a new scenario
for transition to chaos even for weak field strengths \cite{sankar}.

In this paper we explore this model in the context of AM. We show that, our
generalization leads to a larger parametric space, in comparison to the
standard map, for existence of the AM. We demonstrate that when the border
of AM (which we call a ``beach'') exposes a chain of islands of a lower order
resonance, ``stickiness'' of the beach increases significantly leading to
anomalous transport. We also study the signature of such classical phenomena
in the corresponding quantum system. 

\section {Classical System}

\subsection {The Map}

Let us consider a particle confined in an one-dimensional infinite square
well potential $V_0(x)$ of width $2a$ and the Hamiltonian is
\begin{equation}
H_0 = {p^2 \over 2M} + V_0(x) \;\;\; ; \;\;\; 
V_0(x)=\left\{\begin{array}{ll}
0, & \mbox{for $|x|<a$}\\
\infty, & \mbox{for $|x|\;{\geq}\;a$}
\end{array} \right. \, .
\end{equation} 

\n The particle is subjected to a time periodic pulsed field of period $T$. 
The perturbed system is governed by the Hamiltonian
\begin{equation}
H = H_0 + \epsilon \cos\left({2\pi x\over\lambda}\right)
\sum_{n} \delta\left(n-{t\over T}\right) \, ,
\label{ham}
\end{equation}

\n where $\epsilon$ and $\lambda$ are strength and wavelength of the field
respectively; a train of delta functions accomplishes the time periodic pulse.
Kick to kick dynamics of the particle can be described by the map
\begin{eqnarray}
X_{n+1}&=&{(-1)}^{B_n}(X_n + P_n) + {(-1)}^{B_n+1}\mbox{Sgn}(P_n) B_n \, ,
\nonumber \\[4pt]
P_{n+1}&=&{(-1)}^{B_n} P_n + {K\over 2\pi} \sin(2 \pi RX_{n+1}) \, ,
\label{cmap}
\end{eqnarray}

\n where $B_n=\left[\mbox{Sgn}(P_n)(X_n+P_n)+1/2\right]$; $[..]$ and
$\hbox{Sgn}(..)$ stand for integer part and sign of the argument respectively. 
$B_n$ is the number of bounces of the particle between the walls during the
interval between $n$th and $(n+1)$th kick. This dimensionless map is related
to physical variables by the scaling:
\begin{equation}
X_n = {x_n \over 2a}, \; P_n = {p_nT \over 2aM}, \;
K = {2 \epsilon {\pi}^2 T^2 \over aM \lambda}, \; R = {2a \over \lambda}\, , 
\end{equation}

\n where $K$ is effective field strength and $R$ is ratio of two
length scales of the system. Henceforth we refer to the classical
mapping (\ref{cmap}) as the {\it well map}.

The well map can be studied by invoking a {\it Generalized Standard Map}
(GSM) 
\begin{eqnarray}
J_{n+1}&=&J_n+{K\over 2\pi}\sin(2\pi R{\theta}_n) \, , \nonumber \\[4pt]
{\theta}_{n+1}&=&{\theta}_{n}+J_{n+1} \hspace*{0.6cm} (\mbox{mod}\;\;1) \, ,
\label{gsm}
\end{eqnarray}

\n since time reversal of GSM and (\ref{cmap}) are quantitatively related.
GSM is defined on a cylinder $(-\infty ,\infty)\times [{-1/2},{1/2})$ and
the standard map is a special case with $R=1$. A detailed study of GSM
can be found in our earlier work \cite{sankar}. The GSM is highly chaotic for
strong field strength (large $K$). In addition, it exhibits chaotic motion
even for weak field strength (small $K$) depending on the parameter $R$. In
order to study the AM in the well map, we invoke the GSM which is amenable
to a detailed analysis.

\subsection {Stability Analysis}

Periodicity of the GSM in $J$ and $\theta$ (with unit period) implies 
existence of AM. These are located at ($J^{\prime}, {\theta}^{\prime}$):
\begin{equation}
J^{\prime} = m \, , \;\;\;\; 
{K\over2\pi} \sin(2 \pi R {\theta}^{\prime}) = l \, ,
\label{amc}
\end{equation}

\n where $m$ and $l$ are integers. They are also termed as step-$|l|$ AM as
the acceleration is $|l|$ at each iteration. Fixed points belong to the
family of AM with $l=0$. For stable AM, required stability condition is
\begin{equation}
-{4 \over KR} < \cos(2 \pi R {\theta}^{\prime}) < 0 \, ,
\end{equation} 

\n and using (\ref{amc}) this becomes
\begin{equation}
|l| < {K \over 2 \pi} < \sqrt{l^2 + {\left({2 \over \pi R}\right)}^2} \, .
\label{amk}
\end{equation}

\n This inequality can also be rewritten as 
\begin{equation}
0 < R < R_1 \, ,
\label{amr}
\end{equation}
\n where 
\[ R_1 = {2\over\pi} \left\{{\left(K\over 2\pi\right)}^2 
- l^2 \right\}^{-1/2} . \]

For stable AM
\begin{equation}
|{\theta}^{\prime}| = {1 \over 2 R} \left\{2j 
+ 1\mp{1\over\pi}\;{\sin}^{-1}\left({2\pi |l|\over K}\right)\right\} \, ,
\label{theta}
\end{equation}

\n where $j$ is an integer. Thus the AM are characterized by two integers
$l$ and $j$ and we call them as ${(l,j)}_{\mp}$ type modes. Since the 
cylindrical phase space of GSM has the constraint $|{\theta}^{\prime}|\le 1/2$
(equivalent to restriction of the particle dynamics to between walls of the
well), zero can not be lower limit in the above inequality. 

The parameter $R$ in GSM is recognized as frequency of the sinusoidal force
term. As $R$ increases from zero $|{\theta}^{\prime}|$ decreases such that
${(l,j)}_{-}$ type modes first appear in the phase space followed by the
${(l,j)}_{+}$ type. In a similar way, as $R$ decreases $|{\theta}^{\prime}|$
increases such that ${(l,j)}_{+}$ type is the first to disappear from the
phase space while the ${(l,j)}_{-}$ type mode is the last to disappear. Hence,
lowest limit for the inequality (\ref{amr}) is set by the disappearance of
${(l,j)}_{-}$ type modes. From Eqn. (\ref{theta}), the constraint 
$|{\theta}^{\prime}| \le 1/2$ for ${(l,j)}_{-}$ becomes $R_0 \le R$ where
\[ R_0=2j+1-{1\over\pi}\,{\sin}^{-1}\left({2\pi|l|\over K}\right)\, . \] 
Thus the inequality (\ref{amr})
is replaced by
\begin{equation}
R_0 \le R < R_1 \, .
\label{amrr}
\end{equation}

We emphasize that both the inequalities (\ref{amk}) and (\ref{amrr}) must be
simultaneously satisfied for existence of AM in the phase space. This is in
contrast to the standard map for which the inequality (\ref{amk}) with $R=1$
itself is sufficient. Thus the two control parameters, $K$ and $R$, provide a
larger parametric space, in comparison to that of the standard map, for the
existence of AM. At $K=2\pi|l|$, the lower bound of the inequality (\ref{amk}), 
$R_0=1/2$ and this is the minimum possible value of $R_0$. In other words, for
$R<1/2$ AM do not exist. In fact, GSM is hyperbolic for $R<1/2$ and there
are only unstable orbits in the phase space. 

\subsection {Results}

\begin{figure*}[ht]
\centerline{\includegraphics[width=0.9\linewidth]{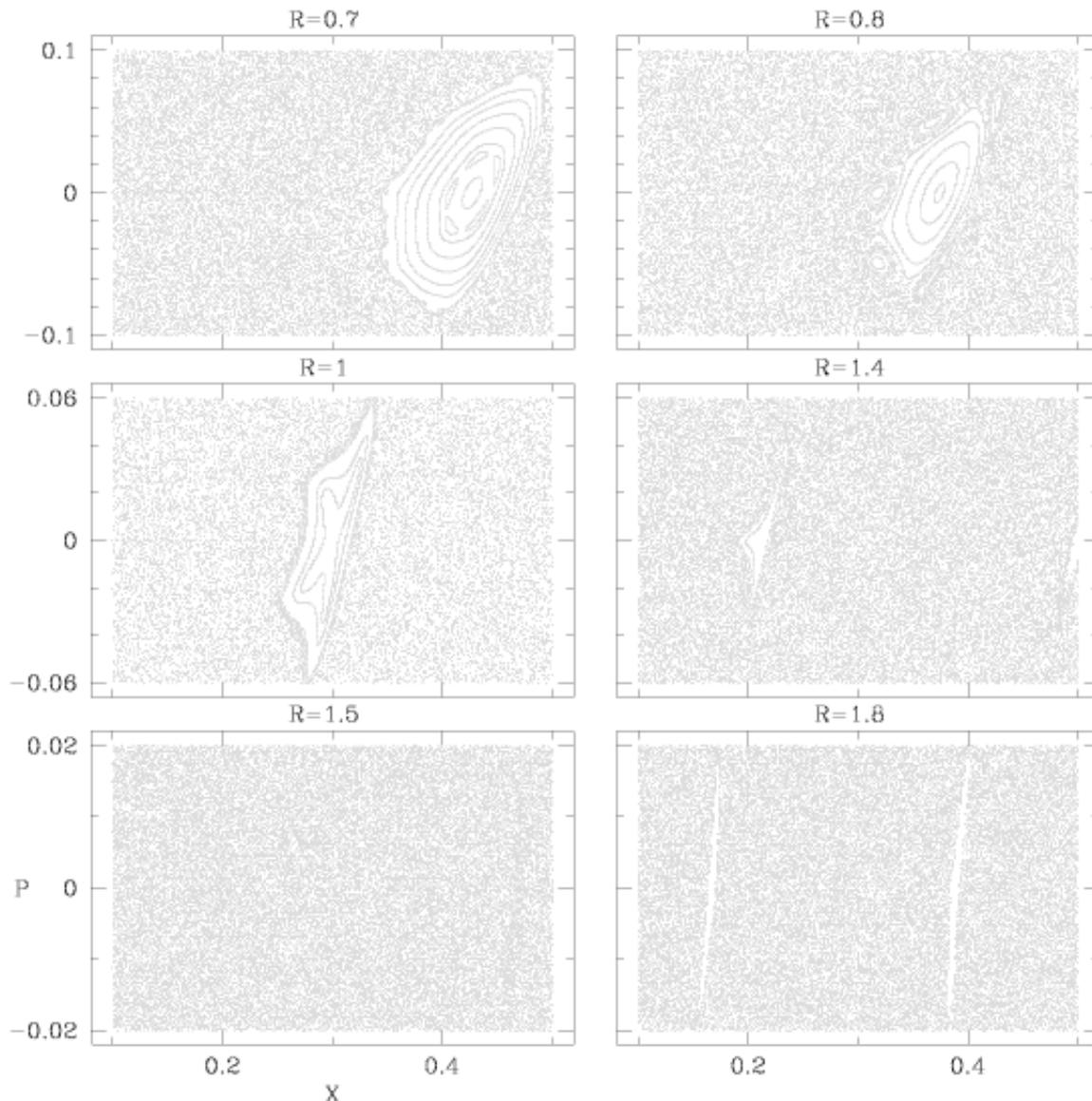}}
\vspace*{-0.8cm}
\caption{Stable regions embedded in chaotic sea are the accelerator modes
of the well map ($K/2\pi=1.05$) whose $P^{\prime}=0$ and $X^{\prime}$ is
positive. For $R=0.7,0.8$ and 1, we observe ${(1,0)}_{-}$ type mode. For 
$R=1.4$, one more mode just appears in the phase space which is ${(1,0)}_{+}$
type. For further increase of $R$, both the types are fully visible. Since
$X^{\prime} \propto 1/R$, as $R$ increases the accelerator islands move
towards interior of the square well. Note the three different scales on
$P$-axis.}
\label{am6}
\end{figure*}

Since the well map and the GSM are quantitatively related, location of AM are
same for both the maps i.e., $(X^{\prime},P^\prime)=(\theta^\prime,J^\prime)$
and all the above arguments hold for well map also. However, AM of the well
map differ only in the following way due to the reflective boundary condition.
When particle is on the AM, $n$th kick causes the particle to undergo $n$
bounces between the walls. If $n$ is odd momentum changes its sign, in turn
sign of the particle position is also flipped.

Phase space of a mixed system generally has many intricate structures. In
particular, boundaries of regular region embedded in chaotic sea have
structures at all scales and all of them are hard to discern. Stable AM 
islands are believed to be separated from the remaining chaotic region by
cantori (partial-barriers). Phase space in the vicinity of the cantori is
sticky as it retains long time correlations. When chaotic orbits explore all
parts of the phase space, there are occasions that they pass through the
cantori. When such an event occurs, chaotic orbits stick to the vicinity of
cantori for long time. Thus they are dragged along the modes ballistically
\cite{shle}. This is a mechanism which is known to enhance momentum transport.
Below we demonstrate that the beaches around the AM island when occupied by
a large resonance of small order are highly sticky. This in turn implies
that the cantori surrounding these must have large gaps to allow larger
penetration of itinerant chaotic orbits.

\begin{figure}[ht]
\centerline{\includegraphics[width=1\linewidth]{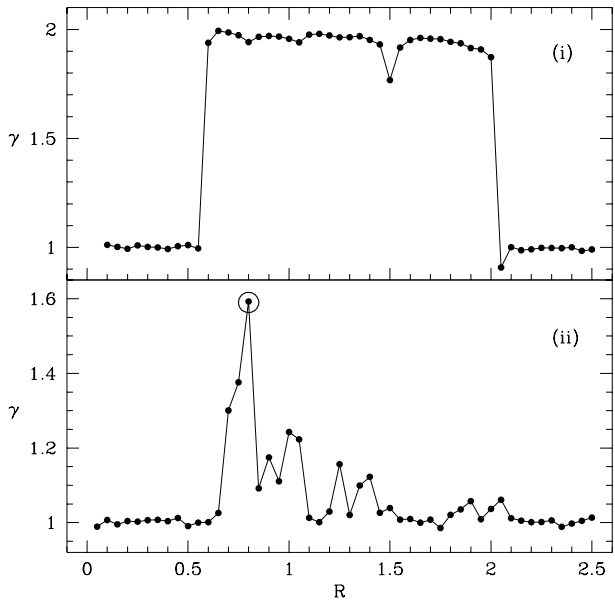}}
\caption{Exponent $\gamma$ is plotted with respect to the parameter $R$ for
$K/2\pi=1.05$. To obtain the exponent, an initial ensemble of 10000 phase
space points is iterated by the well map for 5000 time steps. For (i) the
ensemble consists of points with $P_0=0$ and $X_0$ distributed uniformly
between $-1/2$ and $1/2$. The range of $R$ given by the Eqn. (\ref{amrr})
with $R_0=0.598$ and $R_1=1.988$ is seen as $\gamma\simeq 2$ due to ballistic
evolution of the accelerating points contained in the ensemble. For (ii) the
ensemble is chosen from chaotic region of the phase space. Maximum exponent
observed for $R=0.8$, marked by circle, is due to the beaches around the AM
island.}
\label{gama1}
\end{figure}

Before we dwell more on AM assisted super diffusion, it is instructive to
look at some of the modes of the well map. Fig.~\ref{am6} shows islands for
different $R$ values. In the given range of phase space we find one
${(1,0)}_{-}$ type mode for $R=0.7,0.8$ and 1. As $R$ increases the islands
move towards interior of the well. For $R=1.4$, one more mode just appears
at boundary of the well. This is a ${(1,0)}_{+}$ type mode. For further large
values of $R$ both the types fully appear and they move into the interior
of the well. Evidently, parameter $R$ also changes the size of the islands.

In order to quantify the role of $R$ in AM assisted anomalous transport, 
we consider $\langle {(P_n-P_0)}^2\rangle\sim n^\gamma$. Here the angular 
bracket stands for an ensemble average and the exponent can be evaluated by
iterating an ensemble of phase space points for long time under the well map.
We perform numerical calculations for two different initial ensembles:
(i) phase space points uniformly distributed along the line $P_0=0$ which
correspond to {\it both} accelerator and chaotic orbits; (ii) phase space
points from a chaotic region which include {\it only} chaotic orbits.

The ensemble (i) contains more chaotic orbits than the regular (accelerator)
orbits. Chaotic orbits exhibit normal diffusion in momentum i.e., like the
random walk, until they are dragged along the mode. On the other hand, the
accelerator orbits exhibit ballistic behaviour ($\gamma = 2$). Hence, in
the long time limit, evolution of accelerator orbits dominate the ensemble
average. Fig.~\ref{gama1}(i), shows the characteristic exponent $\gamma$ for
different $R$ values. As we see, for the range of $R$ in Eqn. (\ref{amrr})
$\gamma\simeq 2$, confirming the existence of AM islands. Here the lesser
$\gamma$ for $R=1.5$ is attributed to small size of AM (see Fig.~\ref{am6}).
On contrary, for parameters outside the range of Eqn. (\ref{amrr}), diffusion
is normal as there are no AM in the phase space.

Fig.~\ref{gama1}(ii) corresponds to the ensemble which contains only chaotic
orbits. Within the parametric range where AM exist, the exponent $\gamma$  
shows large fluctuations. There are also occasions wherein presence of AM
does not enhance the diffusion significantly. It is natural to attribute the
observed fluctuations to size of the islands and stickiness of their
neighbourhood. That is, big AM islands with very sticky neighbourhood can
significantly enhance the transport.

For $R=0.8$, the exponent $\gamma$ is found to be maximum ($\simeq 1.6$).
As seen from Fig.~\ref{am6}, in this case AM island is surrounded by a
chain of $1/5$ resonance zones. These zones are also step-1 accelerators
and they are separated from chaotic region by their boundary, the so called
``beaches''. Dynamics on the beach is chaotic, but very sticky with longer
classical staying time. If a wandering chaotic orbit happens to approach
the beach region, it stays there for long time and thus gets accelerated
along the modes resulting in very large enhancement in momentum transport.
This maximum transport is a generic behaviour i.e., independent of choice of
the chaotic ensemble. From Fig.~\ref{am6} we note that, although size
of the AM for $R=0.7$ is larger than that for $R=0.8$, exponent $\gamma$
is maximum for the latter. This shows that the emergence of lower order
resonance zones are chiefly responsible for the stickiness of the beach
regions, leading to AM assisted anomalous transport. Thus the well map or
equivalently GSM has a control parameter $R$ by tuning which AM induced super
diffusion can be maximized. 

More over, longer the period of resonance zones lesser the stickiness of 
associated beach regions. Upon magnifying boundaries of AM islands for
$R=0.7$ and $1$, we found long chain of resonance zones and their beach
regions. As seen from Fig.~\ref{gama1} (ii), enhancement is limited for both
these cases. 

\begin{figure}[ht]
\centerline{\includegraphics[width=1\linewidth]{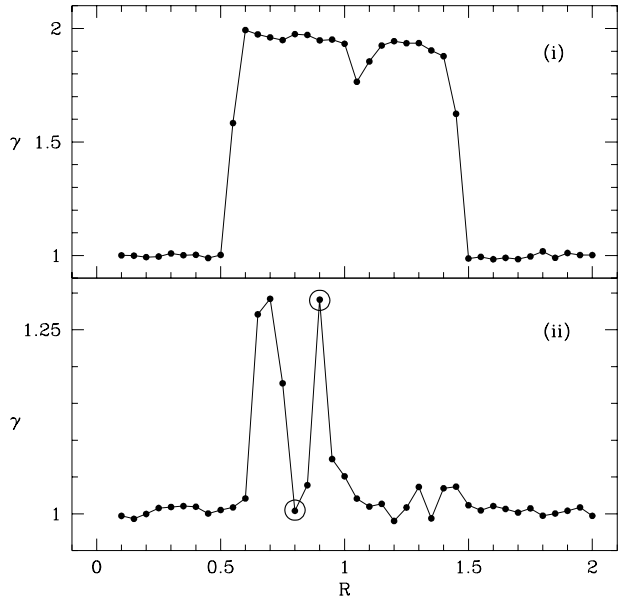}}
\vspace*{-0.3cm}
\caption{Exponent $\gamma$ is plotted with respect to the parameter $R$ for
$K/2\pi=2.05$. Rest of the calculation details are similar to the Fig.~
\ref{gama1}. In (i) the range of $R$ given by the Eqn. (\ref{amrr}) with
$R_0=0.57$ and $R_1=1.41$ is seen as $\gamma\simeq 2$. In (ii) the exponents
for $R=0.8$ and 0.9 are marked by circles. Corresponding AM islands for these
two cases are shown in Fig.~\ref{am2}.} 
\label{gama2}
\end{figure}

Further, we explore step-2 AM which are smaller in size than the step-1 modes.
We choose $K/2\pi=2.05$ and step-2 modes exist in the range $0.57<R<1.41$.
Nearly ballistic evolution in this range and the otherwise normal diffusion
are shown in Fig.~\ref{gama2}(i). In Fig.~\ref{gama2}(ii) the exponent
exhibits large fluctuations. In order to appreciate the role of beach regions
in large scale transport, we focus our attention on two distinct cases where
the exponent shows a minimum and a maximum. 

\begin{figure}[ht]
\centerline{\includegraphics[width=1\linewidth]{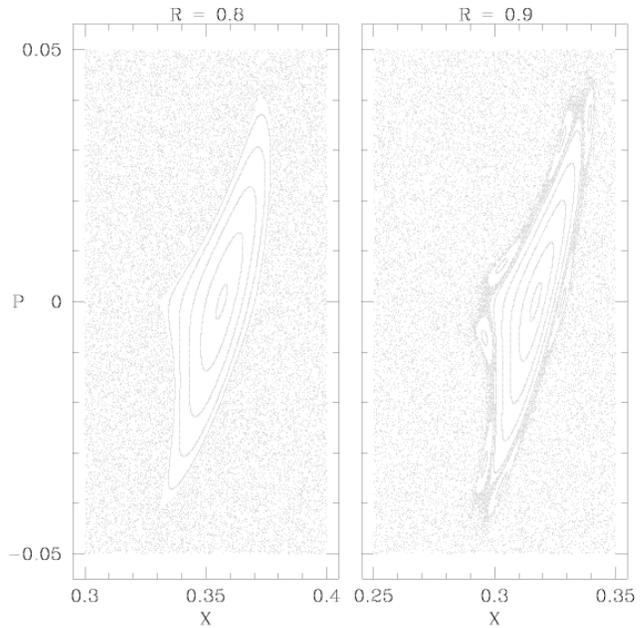}}
\vspace*{-0.3cm}
\caption{Step-2 AM are shown for $K/2\pi=2.05$. For $R=0.9$, the AM island
is accompanied by a chain of $2/7$ resonance zones and sticky beaches. On the
other hand, AM island for $R=0.8$ is not accompanied by any beach regions.}
\label{am2}
\end{figure}

As we see from Fig.~\ref{am2}, when the enhancement is minimum (or absent)
AM islands do not have the surrounding beach regions. On the other hand,
transport is maximum when AM island is surrounded by short chain of resonance
zones and their beach regions. These results reinforce that the expose of
lower order resonance zones around the AM are chiefly responsible for
stickiness of the beach and thus the resulting enhanced momentum transport.

\section {Quantum System}

\subsection {The Map}

Analogous to the classical system, the kick to kick quantum dynamics can be
studied using quantum map by solving Schr\"{o}dinger equation for one period
$T$ as
\begin{equation}
|\psi (t+T)\rangle = U|\psi(t) \rangle \, ,
\label{qmap}
\end{equation}

\n where $U = \exp\{-ik\cos(2\pi x/\lambda)\} \exp\{-iH_0 T/\hbar\}$
with $k=\epsilon T/\hbar$, being the effective field strength. Eigenvalue
equation of the unperturbed system is $H_0|\phi_n\rangle=E_n|\phi_n\rangle$.
The eigenstates and eigenvalues are given by 
\begin{eqnarray}
\langle X|\phi_n\rangle &=& \left\{\begin{array}{ll}
         \sqrt{2} \cos(n\pi X), & \mbox{for $n$ odd}\\[4pt]
         \sqrt{2} \sin(n\pi X), & \mbox{for $n$ even}
                \end{array} \right. \, , \nonumber \\[4pt] 
E_n &=& {n^2{\pi}^2{\hbar}^2 \over 2M} \, ,
\end{eqnarray}

\n with $n=1,2,3\ldots$ Considering unperturbed eigenstates as the basis,
time evolution of an arbitrary quantum state is $|\psi(t)\rangle =
\sum_{n}A_n(t) |\phi_n\rangle$ where $A_n(t) = \langle\phi_n|\psi(t)\rangle$.
With following substitution 
\begin{equation}
\tau = {E_nT \over \hbar n^2} = {{\pi}^2\hbar T \over 2M}
\end{equation}
 
\n quantum map (\ref{qmap}) takes the form
\begin{equation}
A_m(t+T) = \sum_n U_{mn}A_n(t) \, ,
\label{qmap1}
\end{equation}
\n where $U_{mn} = e^{-i\tau n^2}
\langle\phi_m|\exp\{-ik\cos(2\pi RX)\}|\phi_n\rangle$.

Dimensionless quantum parameters $k,\tau$ and the classical parameters are
related through $K/R = 8k\tau$. Since $k\sim 1/\hbar$ and $\tau \sim \hbar$,
semiclassical limit for given classical system can be achieved by taking
the limits $k \rightarrow \infty$ and $\tau \rightarrow 0$. In the following
numerical calculations, quantum map (\ref{qmap1}) is implemented with a
finite, say N, number of unperturbed basis states. By obtaining the limit
\begin{equation}
\lim_{N\rightarrow \infty} \sum_{n=1}^{N}|A_n(t)|^2 \rightarrow 1
\end{equation}

\n we ensure that such a truncation of basis does not influence the quantum
system.

As seen from Eqn. (\ref{qmap1}), unperturbed motion of the particle between
successive kicks just adds phase to the wave function components through the
parameter $\tau$ (mod $2\pi$). If $\tau = 2\pi$, unperturbed motion is absent
and this is called {\it quantum resonance}. Under this condition, without
loss of generality, we can write  
\begin{equation}
|\psi(t)\rangle = e^{-ik\cos(2\pi RX)t}|\psi(0)\rangle \, ,
\label{qres}
\end{equation}

\n where $t$ represents the number of kicks. Then in the limit $t\rightarrow
\infty$, kinetic energy of the particle grows quadratically i.e., ${\langle
E\rangle}_t \sim t^2$ \cite{sankar1}. This phenomenon is similar to that of
the quantum kicked rotor \cite{cas}. An earlier study has shown that quantum
resonance of kicked rotor occurs when $\tau$ is rational multiples of
$2\pi$ \cite{izra}. This non-generic pure quantum phenomenon does not have
any classical analogue. In order to investigate signatures of AM in the
quantum system, it is necessary to suppress the resonance effect. This can
be accomplished by taking $\tau$ as an irrational multiple of $2\pi$.

\subsection {The wave packet}

One natural way to study the quantum signatures of the AM is using the time
evolution of an initial wave packet localized in chaotic region of the
classical phase space. We take a Gaussian wave packet confined in the square
well as the initial state. In position representation it is given by
\begin{equation}
\langle X|\psi(0)\rangle = C \exp\left\{ -{{(X-\langle X\rangle)}^2 
\over 2\sigma^2} + {i\langle P\rangle X \over {\hbar}_e} \right\} \, ,
\label{gaussian}
\end{equation}

\n where $\sigma$ measures width of the wave packet, centred at $\langle X
\rangle$ with momentum $\langle P\rangle$. Here $\hbar_e = 2\tau/\pi^2$ is
the effective Planck constant. Normalization constant, $C$, is obtained from
the condition 

\[ \int_{-1/2}^{1/2}{|\langle X|\psi(0)\rangle|}^2\; dX = 1 \]

\n as
\begin{equation}
C={\left({2/\sigma\sqrt{\pi} \over 
\hbox{erf}(y_+) - \hbox{erf}(y_-)} \right)}^{1 \over 2} \;\; \hbox{with} \;\; 
y_{\pm} = {\pm 1/2 -\langle X\rangle \over \sigma} \, .
\end{equation}  

As the quantum map is described in unperturbed basis states, the initial wave
packet is represented in this basis. For evaluating $A_n(0)$ we consider
following integral
\begin{equation}
G_n = \sqrt{2}\int_{-1/2}^{1/2} e^{in\pi X}\ \langle X|\psi(0)\rangle \ dX \ .
\end{equation}

\n With change of variable $u=(X-\langle X\rangle)/\sqrt{2}\sigma$ the
integral becomes 
\begin{equation}
G_n = 2\sigma C \ e^{iz_n\langle X\rangle} 
\int_{u_-}^{u_+} e^{-(u^2-i\sqrt{2}\sigma z_nu)}\ du \, ,
\end{equation}

\n where $z_n = n\pi+\langle P\rangle/\hbar_e$ and $u_{\pm}=y_{\pm}/\sqrt{2}$.
Using the standard integral \cite{abram}
\begin{eqnarray}
\int e^{-(a_o x^2+2bx+c)}\ dx \hspace*{4cm} \nonumber \\
= {1\over 2}\sqrt{{\pi \over a_o}} \
\exp\left({b^2-ca_o \over a_o}\right) \ \hbox{erf}\left({a_o x + b \over 
\sqrt{a_o}}\right) \, ,
\end{eqnarray}

\n $G_n$ is represented in terms of complex error function, which can be 
computed using the algorithm \cite{gauti}. The wave packet in the unperturbed
basis is then given by 
\begin{equation}
A_n(0) =\left\{\begin{array}{ll}
(G_n + G_{-n})/2, & \mbox{for $n$ odd}\\
(G_n - G_{-n})/2i, & \mbox{for $n$ even}
\end{array} \right.\ . 
\end{equation}

It should noted that because of the spatial confinement, the initial state 
is a truncated Gaussian and hence it is {\it not} a minimum uncertainty wave
packet. However, it can be made close to the minimum uncertainty wave packet
provided mean $\langle X\rangle$ is not close to boundary of the well and
$\Delta X \ll 1$, i.e., spread in position is much less than the width of
the square well. With this condition on the wave packet we consider the
uncertainty relation $\Delta X \Delta P = {\hbar}_e/2$ for the following 
exercise. We take $\Delta X = 0.02$ throughout.

\subsection {Results}

The initial wave packet is chosen with $(\langle X\rangle,\langle P\rangle)
= (0,0.5)$ such that it is placed in a chaotic region. We take ${\hbar}_e
= 0.008$ and hence $\Delta P= 0.2$. For these choice of parameters,
the initial state is squeezed in position and elongated in momentum. The
number of basis states considered is as high as $N=7500$. Dimensionless
kinetic energy of the time evolved state is then given by 
\begin{eqnarray}
{\langle E \rangle}_t &=& \langle\psi(t)|\hat{P}^2|\psi(t)\rangle 
\nonumber \\[6pt]
&=& {\left({2\tau \over \pi}\right)}^2 \sum_{n=1}^{N}{|A_n(t)|}^2n^2 \, , 
\end{eqnarray}

\n which is equivalent to the classical energy $\langle P_t^2\rangle$. 

\begin{figure}[ht]
\centerline{\includegraphics[width=1\linewidth]{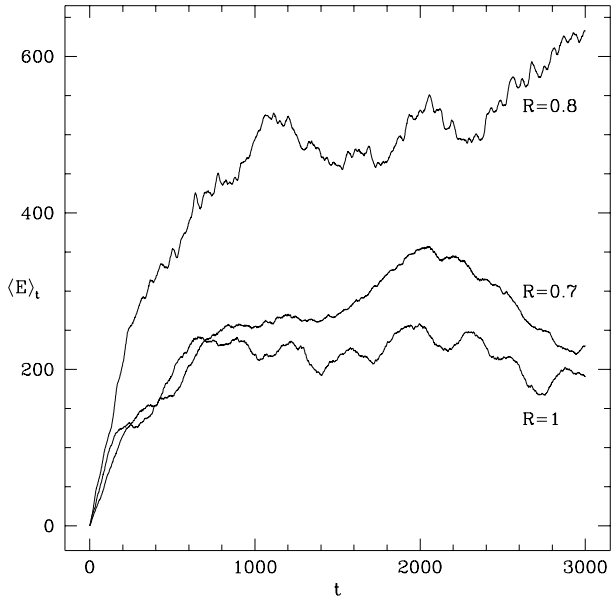}}
\caption{Kinetic energy of the quantum particle for $K/2\pi=1.05$. For
$R=0.7,0.8$ and 1 the effective field strengths are $k=30.3,26.5$ and 21.2
respectively.}
\label{eng_longT}
\end{figure}

Shown in Fig.~\ref{eng_longT} are typical typical quantum evolutions for
$K/2\pi = 1.05$ with varying $R$ values. In all the cases, initially the
quantum energy increases and then attains a quasiperiodic saturation.
However, the saturation is much higher for $R=0.8$ than for $R=0.7$ and 1.
It should be noted that even though effective field strength $k$ is larger
for $R=0.7$ than for $R=0.8$, energy saturation for the later is much higher.
We may compare these results with the corresponding phase space shown in
Fig.~\ref{am6}. Thus we obtain a clear evidence that the expose of lower
order resonance zones and the associated sticky beach around the AM are 
mainly responsible for AM assisted enhanced transport in quantum system
also.

\section {Discussion}

We studied accelerator modes for a particle inside a 1D infinite square well
in presence of a pulsed external field. Influence of modes in altering large
scale transport of the classical system has been studied by invoking an
equivalent dynamical model, the generalized standard map. The virtue of the
generalization arises from the ratio of two length scales, namely, well width
and field wave length. This generalization is found to have larger parametric
space for existence of AM. 

We demonstrated that when the border of AM (beach) exposes short periodic
resonance zones, stickiness of the beach region is significantly increased,
leading to AM assisted super diffusion in momentum. It also appears that 
longer the period of resonance lesser is the stickiness of the beach, results
to less enhancement in transport. The corresponding quantum system is studied
using time evolved wave packet. We showed that the presence of lower order
resonance around the modes are responsible in enhancing the quantum transport
as well.

Some investigations on standard map show that for certain ``magic''
values of system parameter, AM islands are accompanied with hierarchy of
self-similar structures. The vicinity of these structures are known to be
regions of stickiness for the chaotic orbits \cite{zas97,zas97a}, causing
enhancement in transport. Further, various multi-island structures in
the near threshold regime of the map are identified as trapping zones for
wandering orbits \cite{zas98} leading to L\'{e}vy-like flights. All these
results collectively support that island-around-island structures are greatly
responsible for anomalous transport in the mixed phase space. Using fractional
kinectic equation \cite{zas97,zas02}, it will be of interest to enquire the
connection between the period of boundary islands and stickiness of the
associated beach region. \\

\n {\bf Acknowledgement} 

One of us, VBS, would like to acknowledge the hospitality of the Institute
of Theoretical Atomic and Molecular Physics, Harvard University, Cambridge,
USA, during August 2002 when this paper was being completed and written.
We gratefully acknowledge very useful discussions with Dr. A. Lakshminarayan.

\end{document}